\documentclass[twocolumn]{aastex631}
\pdfoutput=1 
\usepackage{amsmath,amstext}
\usepackage[T1]{fontenc}
\usepackage{apjfonts} 
\usepackage[figure,figure*]{hypcap}
\usepackage{multirow}
\usepackage{graphicx}
\usepackage{booktabs}
\graphicspath{{figs/}}

\begin{document}
\title{Studying Postmerger Outflows from Magnetized Neutrino-cooled Accretion Disks} 

\author[0000-0003-1449-0824]{Fatemeh Hossein Nouri}
\author[0000-0002-1622-3036]{Agnieszka Janiuk}
\affiliation{Center for Theoretical Physics, Polish Academy of Sciences,
Al. Lotnik\'ow 32/46, 02-668 Warsaw, Poland}
\author{Małgorzata Przerwa}
\affiliation{Faculty of Physics, University of Warsaw,
Pasteura 5, 02-093 Warsaw, Poland}

\begin{abstract}
    Neutrino-cooled accretion flow around a spinning black hole, produced by a compact binary merger is a promising scenario for jet formation and launching magnetically-driven outflows. Based on GW170817 gravitational wave detection by LIGO and Virgo observatories followed by electromagnetic counterparts, this model can explain the central engine of the short duration gamma ray bursts (GRB) and kilonova radiations. Using the open-source GRMHD HARM-COOL code, we evolved several 2D magnetized accretion disk-black hole models with realistic equation of state in the fixed curved space-time background. 
    We applied particle tracer technique to measure the properties of the outflows.
    The disk and black hole’s initial parameters are chosen in a way to represent different possible post-merger scenarios of the merging compact objects. Our simulations show a strong correlation between black hole’s spin and ejected mass. Generally, mergers producing massive disks and rapidly spinning black holes launch stronger outflows.  We observed our models generate winds with moderate velocity ($v/c \sim 0.1-0.2$), and broad range of electron fraction.  We use these results to estimate the luminosity and light curves of possible radioactively powered transients emitted by such systems. We found the luminosity peaks within the range of $10^{40}-10^{42}$ erg/s which agrees with previous studies for disk wind outflows.
    
\keywords{accretion, accretion disks, gamma-ray bursts, hydrodynamics, magnetohydrodynamics (MHD)}

\end{abstract}

\section{Introduction}\label{intro}

Observation of gravitational waves accompanied by electromagnetic emissions from black hole-neutron star (BHNS) and binary neutron star (BNS) mergers provide treasured information about the physics of the dense matter (\cite{Flanagan2008}), compact binary formation (\cite{Broekgaarden2021}) and the origin of heavy elements (\cite{Kasen2017,Pian2017,Watson-2019}).      

In a post-merger black hole-accretion disk remnants, energy can be channeled into ultra-relativistic outflow needed to explain GRB properties(~\cite{Nakar2007,Berger:2013jza}) by hot, dense accretion flow.
In such systems, the plasma cools by neutrino emission continuously (\cite{Popham99,DiMatteo2002,Janiuk-2004}). On the other hand, magnetic fields can also extract energy from the disk and black hole spin, by generating the magnetically-driven winds and Poynting flux-dominated jets (\cite{BZ77}). Moreover, it can support the heating process of the plasma through the viscous effects of the MRI (\cite{BalbusHaw1991}). 
The neutron-rich dynamical ejecta outflows during merger, and post-merger winds are considered as possible sources explaining the observation of kilonovae, afterglows and other electromagnetic (EM) counterparts following the gravitational wave detections, (\cite{Metzger:2010,metzger:11,Dietrich-2017,Zhu-2021}). 

Kilonovae are transient emissions in the optical or near-infrared band,
powered by the radioactive decay of the elements produced
through r-process nucleosynthesis (\cite{Tanvir-2013}).
Two distinguishable sources are suggested to explain kilonova emissions:
1- High-speed, neutron-rich dynamical ejecta expelled from the tidal tail during merger; 
2- Less neutron-rich material with moderate velocity ejected from post-merger accretion disk due to magnetic viscous heating and/or neutrino absorption heating effects.
The emissions from former source are dimmer and long-lasting near infrared (\cite{Li-Paczynski-1998,metzger:11,Grossman-2014,Tanaka-2014}), 
while the latter produces optically brighter and bluer transients 
(\cite{Grossman-2014,Perego-2014,Kasen-2015}). 
However, depending on the merger's scenario, it is possible to have multiple channels for outflow formation; in more recent studies by~\cite{combi-2022} it is shown that the decay of the fast free neutrons in the outermost layers of dynamical ejecta can power a UV/optical kilonova precursor on $\approx 1$ hour  timescale.
In the case of a BNS merger, the long-lived postmerger hypermassive neutron star (HMNS) is considered as another channel of neutrino and magnetically-driven outflow formations (~\cite{Curtis-2021,Haas-2022}). 

The observation of the electromagnetic counterpart of the gravitational wave signal from the BNS merger GW170817 gives us a unique opportunity to investigate the r-process nucleosynthesis directly. 
The identification of strontium in these emissions (\cite{Watson-2019}) is considered as the first evidence of rapid nuclear process and heavy elements production in the BNS merger scenario. 
The observed kilonova light-curves and spectra suggest that
models with two components consistent with lanthanide-poor and lanthanide-rich ejecta provide a good fit to the data. 
These models indicate the ejecta mass of $M_{ej}/M_{\odot} = 0.03-0.06$, with postmerger disk ejeta as a dominant component (\cite{Cowperthwaite:2017,Tanaka:2017,Kawaguchi:2018}).
The theoretical models suggest that the magnitude, color, and duration of a kilonova are significantly affected by the mass, velocity, and composition of the ejected matter, and the properties of these outflows can themselves be related to the properties of the merging compact objects, such as mass ratio, compactness and equation of state (\cite{Barnes-2013,Grossman-2014}).

In a post-merger scenario, the jet formation, disk evolution and ejected matter can be impacted by the magnetic field strength and geometry. ~\cite{Wan-2017} studied the long evolution of a BHNS merger where the neutron star is initially magnetized by asymmetric magnetic field dipole configurations. 
~\cite{Christie-2019} evolved a long-term ($\sim 4$s) post-merger black hole-accretion disk system with different magnetic field configurations, concluding the ejected matter is slightly more massive for poloidal field configuration compared with purely toroidal field.
~\cite{Janiuk-2019} also reported that disk configurations with higher initial magnetic field strength and higher spin BH generate massive, lanthanide-poor outflows. 
Moreover, the total mass lost through wind outflows depends on the mass of the remnant torus as well. Larger and more massive disks generates stronger matter outflows (\cite{Kasen2017}).
However, only a long-term 3D GRMHD simulations of a posmerger system can capture all the disk's and outflows features accurately, including the geometry and thermal evolution. 
~\cite{Fernandez-2019} and~\cite{Hayashi-2021} have evolved multi-second GRMHD simulations of black hole-disk remnants for BNS and BHNS mergers respectively. Both studies concluded that the neutrino cooling is only effective at the earlier time after merger, and the disk eventually becomes advective, hence the mass loss rate and ejecta's composition are expected to be affected by long evolution of the disk.  

Predicting the properties of matter outflows in BNS and BHNS mergers is crucial to carry out detailed analysis of post-merger electromagnetic emissions.
~\cite{Li-Paczynski-1998} and ~\cite{Grossman-2014} obtained the analytical fits for the mass, kinetic energy, and the velocities of the dynamical ejected material to approximate the main properties of kilonovae according to BNS numerical simulations. 
~\cite{Dietrich-2017} suggested fitting formulae to derive the dynamical ejecta's characteristics based on the results from over 170 BNS merger numerical simulations.
Most recently,~\cite{Holmbeck-2021} introduced a novel method on estimating the individual masses and equation of state of BNS systems from r-process abundance signatures and numerical simulations using Bayesian analysis. 
~\cite{Raaijmakers-2021} is another recent paper in the Bayesian framework, which improves the estimation of the gravitational wave source parameters using the kilonova's lightcurves in the case of BHNS systems. 

In order to investigate the properties of outflows, it is important to have a reliable method to determine the unbound matter and its evolution.
The accurate measurement of the outflow's quantities and geometry can be captured only by 3D hydrodynamics simulations including the r-process heating and neutrino cooling terms in evolution equations (\cite{Klion:2022,Foucart2021}). However, ~\cite{Desai-2019} and ~\cite{Foucart2021} presented improved versions of Bernoulli's criteria to
include the effects of these heating and cooling process approximately, to have a realistic estimation of
the unbound matter.

In this paper, we utilize the numerical simulations to investigate the evolution and different properties of the disk and the outflows in the presence of magnetic fields and neutrino cooling. 
We mainly focus on measuring the outflows’ properties such as composition, velocity and mass to estimate the kilonova’s luminosity and lightcurve. We also investigate how the initial analytic disk parameters can affect the features of the post-merger outflows by altering disk's mass and black hole's spin and mass. The range of parameters are chosen to mimic real post-merger remnant disks. 

This paper is organized as follows, in Sec.~\ref{sec:numerics-and-setup} we briefly explain the numerical framework and initial setup of our simulations. In Sec.~\ref{sec:results} we present the general properties of the outflows and predict the possible kilonova lightcurves powered by each model. We give a discussion over comparison of our results with the previous studies in the literature and also physical and numerical limits of our study in Sec.~\ref{sec:discussion}. The summary and conclusion is given in Sec.~\ref{sec:conclusion}, and finally, in the appendix~\ref{sec:ejecta-measurment} we present the results from several tests examining the accuracy of our tracer method for outflow measurements.

\section{Numerical Methods and Initial Setup}
\label{sec:numerics-and-setup}

\subsection{HARM code}

We use a developed version of the general relativistic magnetohydrodynamic (GRMHD) code HARM (\cite{Gammie2003}, \cite{Noble2006}) explained in~\cite{Sap2019}.
This version of HARM has been developed to include realistic equation of state as well as neutrino treatment. HARM is a finite-volume code with HLL shock capturing scheme, which solves the partial differential GRMHD equations in the standard Valencia conservation formalism~\cite{Papadopoulos1999}, for continuity equation, energy and momentum conservation,

\begin{equation}
\begin{aligned}
    \left( \rho u_{\mu} \right)_{;\mu} =0 \\
    T^{\mu}_{\nu ; \mu} = 0,
\end{aligned}
\end{equation}

and induction equation,

\begin{equation}
\begin{aligned}
    \partial_t \left( \sqrt{-g} B^i \right) = -\partial_j \left(\sqrt{-g} \left(b^j u^i - b^i u^j\right)\right).
\end{aligned}    
\end{equation}

Here $T$ is the energy-momentum tensor with both matter and electromagnetic contributions $T^{\mu \nu} = T^{\mu \nu}_{gas} + T^{\mu \nu}_{EM}$, $u_\mu$ is the four-velocity, $B$ is the magnetic field and $b^i \equiv (B^i + B^i u^{\mu} g_{i \mu}u^i)/u^t$. 
The metric is frozen and fixed to the Kerr metric. The hydro equations are evolved in the modified spherical Kerr-Schild coordinates, where the following radial and angular maps are applied to decrease the grid spacing close to the black hole and the equatorial plane, respectively to improve the accuracy. 

\begin{equation}
\begin{aligned}
   & r_{KS} = R_0 + e^{r_{MKS}} \\
   & \theta = \pi x^{[2]} + \frac{(1-h)}{2}sin(2\pi x^{[2]})
\end{aligned}
\end{equation}

The coordinate parameter $h$ is set to $0.3$ for all models in this paper.

\subsection{Initial Setup}

This work can be considered as a follow-up studies of~\cite{Janiuk-2019} including a wider range of parameters for disk setup. We choose the initial disk mass in the range of [0.05,0.3]$M_{\odot}$ to match with a realist post-merger remnant disks from a black hole-neutron star (BHNS) or neutron star binary system (BNS)~\cite{Kruger-2020}. In order to investigate the spin effects on the disk ejecta, we alter the spin of the black hole (BH) for a few cases of prograde and one retrograde case. 
The mass of the BH has been set to $\sim 2.65M_{\odot}$ for BNS case, and $~5-6M_{\odot}$ for the BHNS cases. 
In addition, we have a few cases of light black holes (LBH) $1-2M_{\odot}$, which are predicted to be formed through the collisions of low mass primordial black holes with galactic neutron stars in galactic dark matter halos~\cite{Abramowicz-2022}.
The details of our different models are given in table~\ref{tab:models}.

The initial state of the torus is derived from the analytic solution of the Fisbone-Mocrief (FM)~\cite{FM1976} disk model around a Kerr black hole. The FM disk is defined by $r_{in}$ the inner radius of the disk, $r_{max}$ the radius of maximum pressure, and $a$ dimensionless spin of the black hole. The FM disk free parameters in geometrical units, and the equivalent constant specific angular momentum $l_{FM}$ for each model are given in table~\ref{tab:models}. 

The initial magnetic field is confined within the torus with pure poloidal configuration for all the cases defined by vector potential:

\begin{equation}
    A_{\phi} = \frac{\overline{\rho}}{\rho_{max}} - 0.2,
\end{equation}

where $\overline{\rho}$ is the density averaged over density at grid point and the neighboring cells, and $\rho_{max}$ is the maximum density. The strength of the magnetic field is set to $\beta=50$ parameter, where $\beta$ is defined as the ratio of the gas pressure to the magnetic pressure $\beta \equiv P_{g}/P_{B}$.
The resolution of the grid is 384x300 for radial and polar angular direction, respectively for all the models.

\subsection{Equation of State and Neutrino Treatment}

For the nucleus equation of state we follow the same approach as~\cite{Janiuk-2019} to have the plasma composed of free protons, neutrons, electron–positron pairs, and helium nuclei. 
The fraction of each species is determined by the equilibrium condition assumed between the reactions of electron–positron capture on nucleons, and neutron decays based on prescription given by~\cite{Reddy-1998}. 
The gas is in equilibrium, so that the ratio of protons to neutrons satisfies the balance between forward
and backward of the following weak nuclear reactions:
$p + e^- \to n + \nu_e$,
$p + \overline{\nu}_e \to n + e^+$, 
$p + e^- + \overline{\nu}_e \to n$,
$n + e^+ \to p + \overline{\nu}_e$, 
$n + \nu_e \to p + e^-$, and
$n \to p + e^- + \overline{\nu}_e$.

We can define the proton-to-baryon number density ratio $Y_p$
(equivalently the electron fraction, for charge
neutrality condition) as: $Y_e = (n_{e^-} - n_{e^+})/n_b$, where $n_b$ is the baryon number density, $n_{e^-}$ and $n_{e^+}$ are electron and positron number densities respectively.
In our simulations $Y_e$ is not evolved by time but it determines after each time step by the equilibrium conditions. This quantity is used to study the composition of the outflow winds. 
Based on the discussion in~\cite{Woosley-1992} and~\cite{Kasen-2015} the r-process nucleosynthesis is only effective when $Y_e < 0.4$, and the lanthanide elements are more likely to be produced when $Y_e < 0.25$ in the ejected matter. Therefore the accuracy of $Y_e < 0.25$ measurement is crucial to predict the kilonova's features.  

For the neutrino treatment, we applied the same approach as~\cite{Janiuk-2019} where the absorption optical depth for different neutrino species is given by an approximation (\cite{DiMatteo2002}),

\begin{equation}
   \tau_{a,\nu_i} = \frac{H}{4 \frac{7}{8} \sigma T^4} q_{a,\nu_i},
\end{equation}

where $q_{a,\nu_i}$ is the absorption rate derived by summation over the absorption rates from the weak interactions mentioned above, $T$ is temperature, $\sigma$ is the Stefan-Bolzmann constant, and $H$ is disk's height.
The lepton flavour considered here for neutrino species are electron, $\mu$ and $\tau$.  
The scattering optical depth is estimated as:

\begin{equation}
\begin{aligned}
    & \tau_{s} = \tau_{s,p} + \tau_{s,n} \\
    &   = 24.28 \times 10^{-5} \left[ \left(\frac{kT}{m_e c^2})\right)^2 H \left( C_{s,p}n_p + C_{s,n}n_n \right)\right],
\end{aligned}
\end{equation}
  
where $C_{s,p}=[4(C_V-1)^2+5\alpha^2]/24, C_{s,n}=(1+5 \alpha^2)/24, C_V=1/2+2~sin^2 \theta_C$, with $\alpha=1.25$ and $sin^2 \theta_C = 0.23$

The neutrino cooling rate is computed as:

\begin{equation}
    Q_{\nu} = \frac{(7/8) \sigma T^4}{(3/4)} \sum_{i=e,\mu,\tau}\frac{1}{0.5(\tau_{a,i}+\tau_{s}) + 1/\sqrt{3} +1/(3\tau_{\tau_{a,i}})}
\end{equation}

The detailed discussions over nuclear equation of state and neutrino treatment implemented in HARM-COOL code are given in~\cite{Janiuk:2007,Janiuk-2013,Janiuk-2019}.

\begin{table*}
\begin{ruledtabular}
\begin{tabular}{ l l l l l l l l l l}
Model & BH Mass [$M_{\odot}$] & BH Spin & $M_{disk}$ [$M_{\odot}$] & $r_{in}$ & $r_{max}$ & $l_{FM}$ & mass ratio \\
\hline
M1.0-0.14-a0.98 & 1.0 & 0.98 & 0.14062 & 6 & 12 & 4.293 & 0.14\\
M1.0-0.14-a0.9  & 1.0 & 0.9 & 0.14062 & 6 & 12 & 4.293 & 0.14\\
M1.0-0.14-a0.6  & 1.0 & 0.6 & 0.14062 & 6 & 12 & 4.293 & 0.14 \\
M1.0-0.14-a0.2  & 1.0 & 0.2 & 0.14062 & 6 & 12 & 4.293 & 0.14\\
M5.0-0.3-a0.9 & 5.0 & 0.9 & 0.3120 & 6.5 & 13.4 & 4.44 & 0.06\\
M1.5-0.1-a0.9 & 1.5 & 0.9 & 0.09722 & 5.4 & 11 & 4.189 & 0.0635\\
M2.0-0.05-a0.9 & 2.0 & 0.9 & 0.04548 & 4.8 & 9.75 & 4.0617 & 0.0225\\
M2.65-0.1-a0.9 & 2.65 & 0.9 & 0.10276 & 3.8 & 9.75 & 4.0617 & 0.0265 \\
M6.0-0.14-aR0.6 & 6.0 & -0.6 & 0.14213 & 7.8 & 16.8 & 5.148 & 0.024 \\
M6.0-0.14-a0.6 & 6.0 & 0.6 & 0.14062 & 6 & 12 & 4.293 &  0.024 \\
\end{tabular}
\end{ruledtabular}
\caption{\label{tab:models}
Different disk setup for numerical simulations. 
}
\end{table*}

\section{Results} 
\label{sec:results}

\subsection{Outflows: Mass, Composition, Velocity and Geometry}

Measuring the general properties of the outflows such as mass, velocity and composition, with high accuracy is a crucial task to predict the observable kilonovae lightcurves. Using the particle tracer technique would allow us to learn about the geometry of the outflows as well. In this section, we present our quantitative measurements for outflow properties using the tracer method, focusing on the BH spin effects and also BH and disk masses' effects on these properties. At the end of this section, we use these results to estimate the time, luminosity and temperature peaks, and lightcurves of the kilonova emitting from each model.  

The total outflow mass, average velocity and electron fraction, measured by tracers at $r=800 r_g$, are given in Table~\ref{tab:peaks}.
Figs~\ref{fig:massdistr-v-spin}-\ref{fig:massdistr-theta-mass} show mass distribution histograms versus velocity, electron fraction and polar angle $\theta$ for models with different spin and mass configurations. 
A quick observation from these quantities show that all the magnetized-neutrino cooled disk models generate ejecta with moderate velocity $v \sim 0.1-0.2$c, and high electron fraction $Y_e > 0.25$ for regular BNS and BHNS mass configurations, which perfectly describes the second channel of the kilonova emission sources from the postmerger torus producing bright and blue transients in a few hours after merger. The results for LBH cases are different regarding the composition which is discussed in Sec.~\ref{sec:spin-effect} and~\ref{sec:mass-effect}.

\subsubsection{Spin effects}
\label{sec:spin-effect}

Focusing on more details in our results show that the remnant BH spin plays an important role in outflows features.
The models labeled as M1.0-0.14-a0.98, M1.0-0.14-a0.9, M1.0-0.14-a0.6 and  M1.0-0.14-a0.2 all have identical initial parameters except for BH spin.
As shown in Table~\ref{tab:peaks} the mass of ejecta is hugely affected by the BH spin. The mass increases monotonically by the BH spin and the difference may reach to more than three orders of magnitude between the low-spin case M1.0-0.14-a0.2 and extremely high-spin case M1.0-0.14-a0.98.
This is consistent with previous GRMHD numerical studies observing that high spin BH cases with robust magnetically collimated jets produce more massive outflows (\cite{Hawley-2006}). Similar observation about massive ejecta has been reported by~\cite{Fernandez-2015} for viscous accretion disks with highly spinning black holes.  
The significant mass-loss is due to energy release by accretion happening deeper in the gravitational potential and closer to the BH singularity.
This conclusion may let us rule out the merger scenarios with zero or low-spin remnants from the parameter estimation for bright kilonova observations.     

\begin{figure*}
    \centering
    \includegraphics[width=\textwidth]{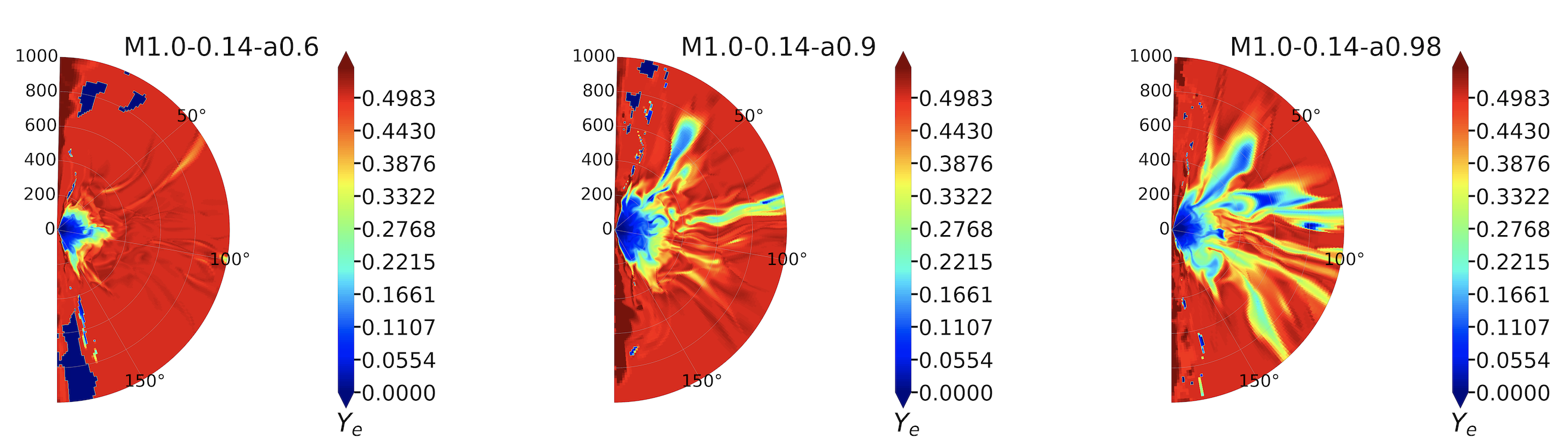}
    \caption{Electron fraction profiles for models with different BH spins at final time snapshot ($t \sim 0.247$s).}
    \label{fig:Ye-spin-2D}
\end{figure*}

Moreover, the 2D profiles of the electron fraction given in Fig.~\ref{fig:Ye-spin-2D} and the average $Y_e$ in Table.~\ref{tab:peaks}, as well as the histograms in Fig.~\ref{fig:massdistr-ye-spin} derived from SkyNet code for r-process nucleosynthesis for models with LBH ($M_{BH} = 1.0 M_{\odot}$) and different spins show a dramatic change in the outflow compositions. Generally, the outflows are more neutron-rich for models with higher BH spin (with average $Y_e < 0.25$), which leads to more opaque lanthanide-rich material. This observation is in contrast with~\cite{Janiuk-2019} and~\cite{Fernandez-2015}, which both found that the electron fraction will increase as the BH spin increases for stellar mass BH. The argument for their observation is, when the accretion happens in a deeper gravitational potential, it releases more energy and heat the inner part of the disk, therefore the weak interactions' equilibrium condition change in a way to release more neutrinos for more effective neutrino cooling. This causes producing more proton-rich plasma. However, in our equation of state, the partial trapping of neutrinos is taken into account (see the details in~\cite{Janiuk-2019} for the calculation of trapped neutrino pressure), so the weak interaction rates are affected by neutrino trapping. In fact, a closer look into some quantities such as neutrino luminosity and neutrino emissivity show that the highest spin case, M1.0-0.14-a0.98, goes through a less effective neutrino cooling during the evolution compared with M1.0-0.14-a0.9 case. 
However, more cases with LBH and stellar mass BH configurations with different spins are needed in future studies to investigate this effect more closely.  

Similarly for velocity, we observe significant changes imposed by BH spin.  Fig.~\ref{fig:massdistr-v-spin} shows the high spin cases such as M1.0-0.14-a0.98 and M1.0-0.14-a0.9 generate outflows with higher and broader range of velocities. However, the average velocity does not increase with BH spin monotonically as our highest spin case M1.0-0.14-a0.98 has the average velocity slightly lower than the M1.0-0.14-a0.9 case.  

From the geometrical point of view, as shown in Fig.~\ref{fig:massdistr-theta-spin} all cases with different spins have very broad range of ejected angles. Almost all the cases have a distinguishable peak around the equator $\theta \sim 80^{\circ}-100^{\circ}$, however it is interesting to mention that the symmetry is somehow broken for some cases, i.e. significantly more wind ejected from the northern hemisphere than the southern hemisphere and vise versa. This geometrical pattern is almost visible for all the cases regardless of their BH spins.

\subsubsection{Prograde versus retrograde}

For a complete comparison, we perform similar analysis between one prograde case with moderate spin $a=0.6$ labeled as M6.0-0.14-a0.6 and one retrograde case with spin $a=-0.6$ labeled as M6.0-0.14-aR0.6. Both cases have BH mass equals to $6M_{\odot}$ and disk mass equals to $0.14M_{\odot}$, however the initial disk parameters such the inner radius and the constant specific angular momentum for each case is different for each model based on the nature of the Fishbone-Mocrief solution (see ~\cite{Kozlowski-1978} for possible retrograde disk solutions). So, the disks are not identical in term of size and compactness. 

The comparison between these two models show that the retrograde case generates much lower mass (by one order of magnitude), and faster ejecta. However the higher velocity can be explained by that fact this disk is initially larger than the prograde case (larger initial specific FM angular momentum) and therefore less bounded to the BH, and produce higher speed ejecta.
On the other hand, as shown in Figs.~\ref{fig:massdistr-v-spin}-\ref{fig:massdistr-theta-spin} the retrograde case has a narrow range of velocities and more symmetric geometry with more outflows leaving the grid around the equator compared with the prograde case. 
Overall, one may conclude that the negative spin can provide distinguishable changes in the amount of the outflows, its velocity and geometry.  

\begin{figure*}
    \centering
    \includegraphics[width=\textwidth]{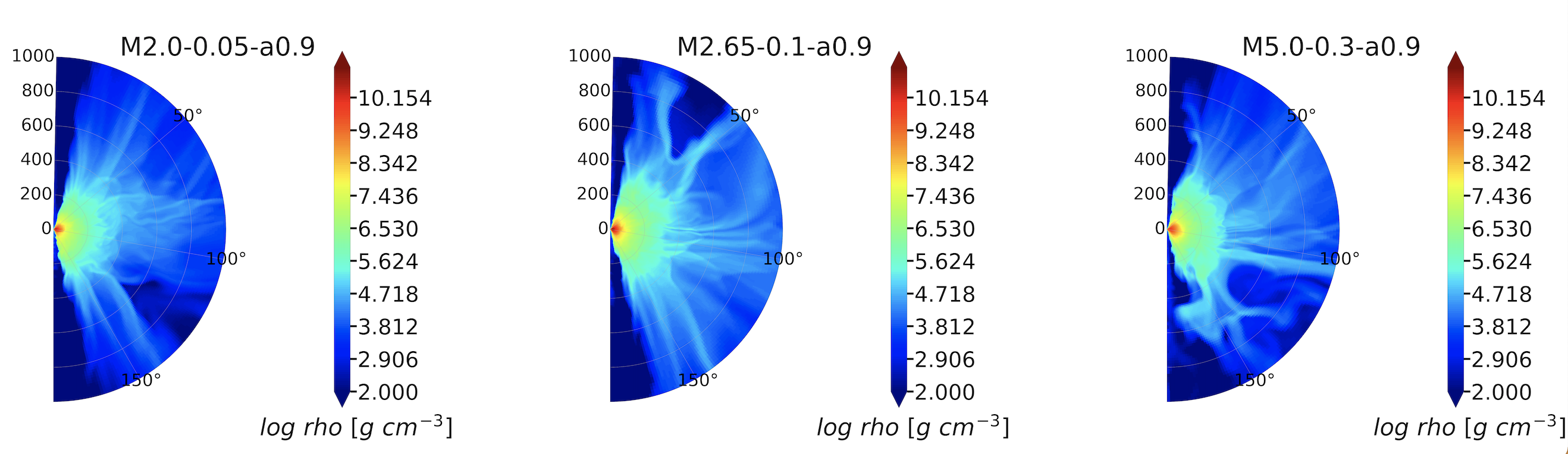}
    \caption{Density profiles for models with different BH mass and disk mass at final time snapshot.}
    \label{fig:Rho-mass-2D}
\end{figure*}

\begin{figure}
\centering
\begin{minipage}{0.48\textwidth}
\includegraphics[width=\textwidth]{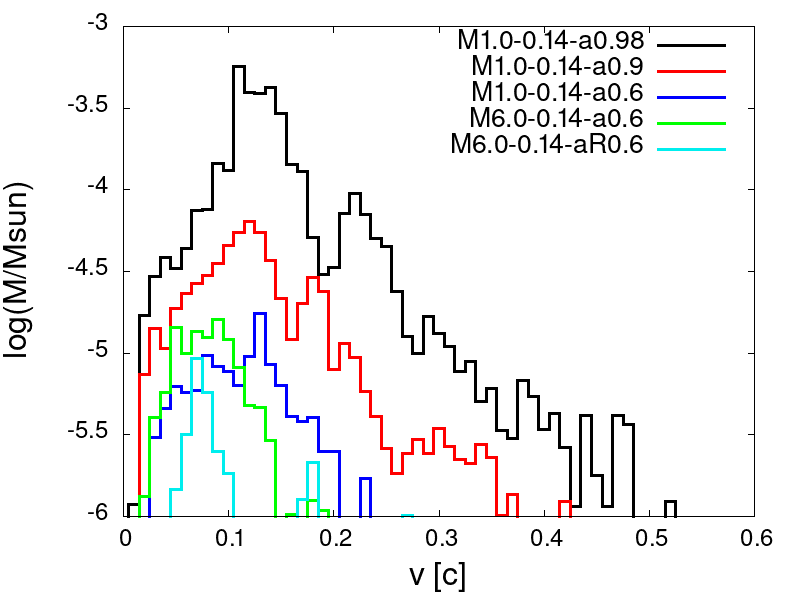}
\caption{The mass distribution versus velocity, comparing cases with different spins for LBH with mass equals to $1M_{\odot}$ and one pair of retrograde and prograde case with BH mass equals to $6M_{\odot}$.}
\label{fig:massdistr-v-spin}
\end{minipage}
\quad
\begin{minipage}{0.48\textwidth}
\includegraphics[width=\textwidth]{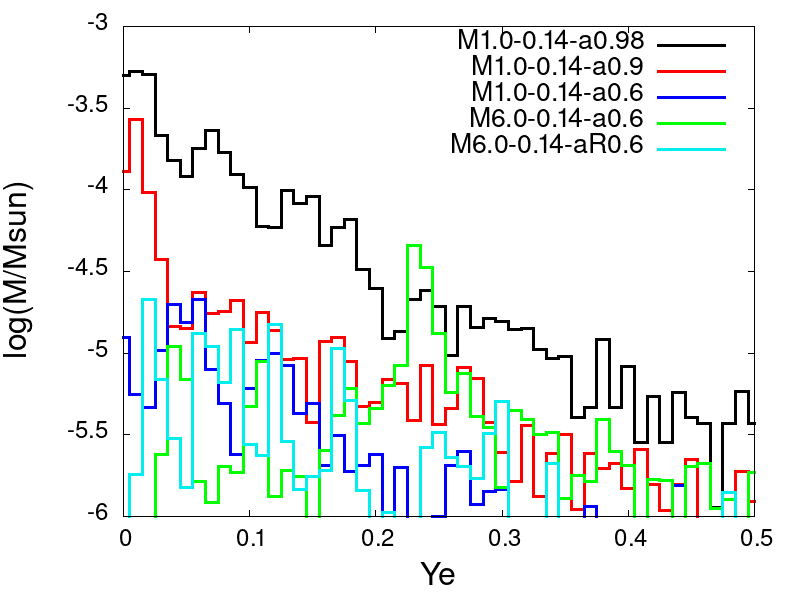}
\caption{The mass distribution versus electron fraction, comparison for cases with spin and mass configurations as Fig.~\ref{fig:massdistr-v-spin}.}
\label{fig:massdistr-ye-spin}
\end{minipage}
\quad
\begin{minipage}{0.48\textwidth}
\includegraphics[width=\textwidth]{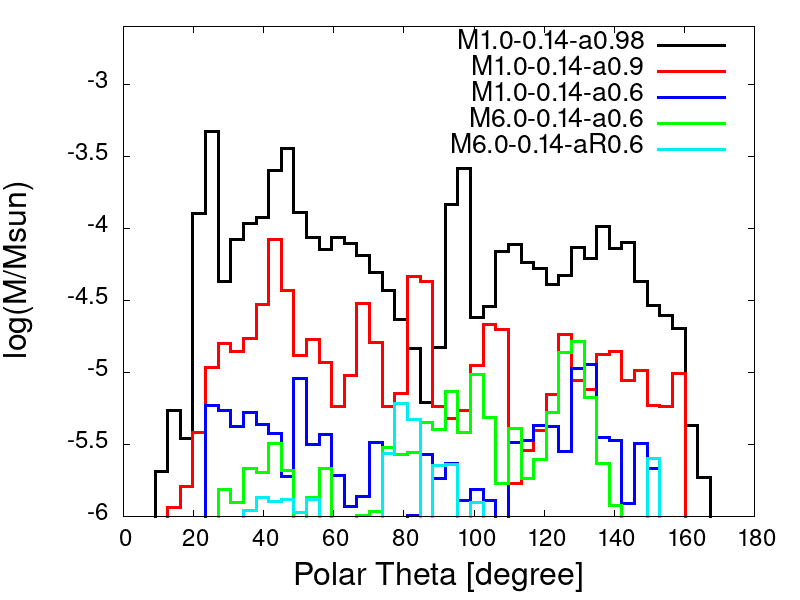}
\caption{The mass distribution versus polar angle, comparison for cases with spin and mass configurations as Fig.~\ref{fig:massdistr-v-spin}.}
\label{fig:massdistr-theta-spin}
\end{minipage}
\end{figure}

\subsubsection{Effects of BH and Disk Masses}
\label{sec:mass-effect}

We found out the postmerger remnant disk's mass has a significant impact on the outflow properties as well. However the changes are less dramatic compared to the spin effects.
The outflow becomes more massive by more than one order of magnitude as move from M2.0-0.05-0.9 case with $M_{disk} = 0.05 M_{\odot}$ to the highest disk mass case M5.0-0.3-0.9 with $M_{disk} = 0.3 M_{\odot}$.
This obviously suggests that mergers with massive remnants is more likely to have observable EM counterpart. Such scenarios are possible from BNS mergers and BHNS mergers with low mass ratio, softer neutron star's equation of state and high spin BH. 

\begin{figure}
\centering
\begin{minipage}{0.48\textwidth}
\includegraphics[width=\textwidth]{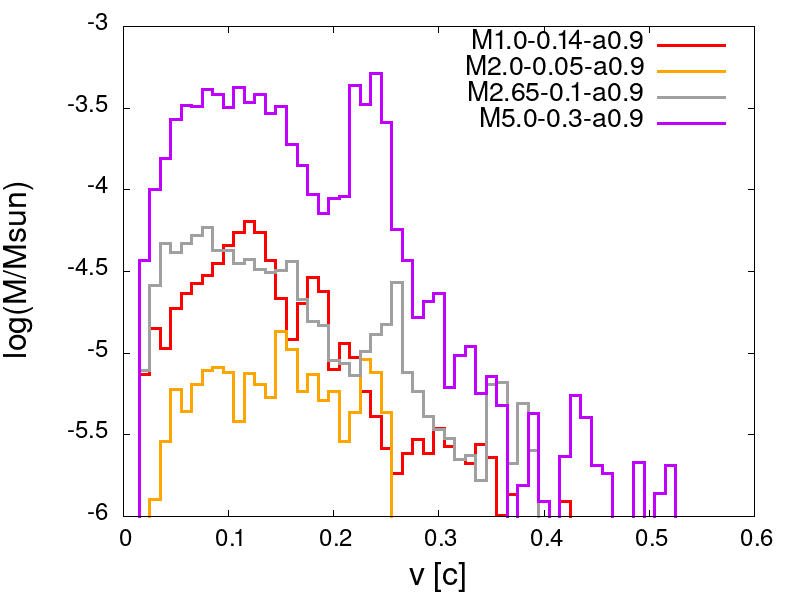}
\caption{The mass distribution versus velocity, comparing cases with different disk and black hole mass configurations.}
\label{fig:massdistr-v-mass}
\end{minipage}
\quad
\begin{minipage}{0.48\textwidth}
\includegraphics[width=\textwidth]{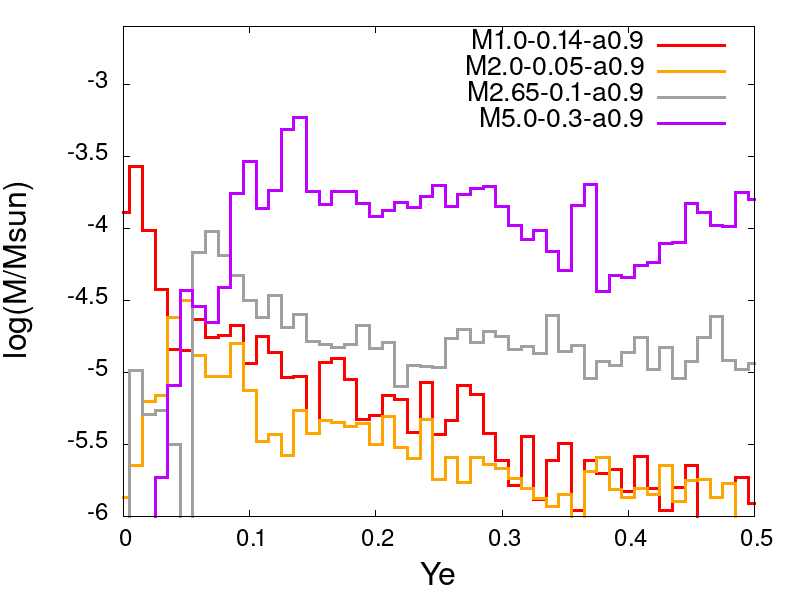}
\caption{The mass distribution versus electron fraction, comparing cases with different disk and black hole mass configurations.}
\label{fig:massdistr-ye-mass}
\end{minipage}
\quad
\begin{minipage}{0.48\textwidth}
\includegraphics[width=\textwidth]{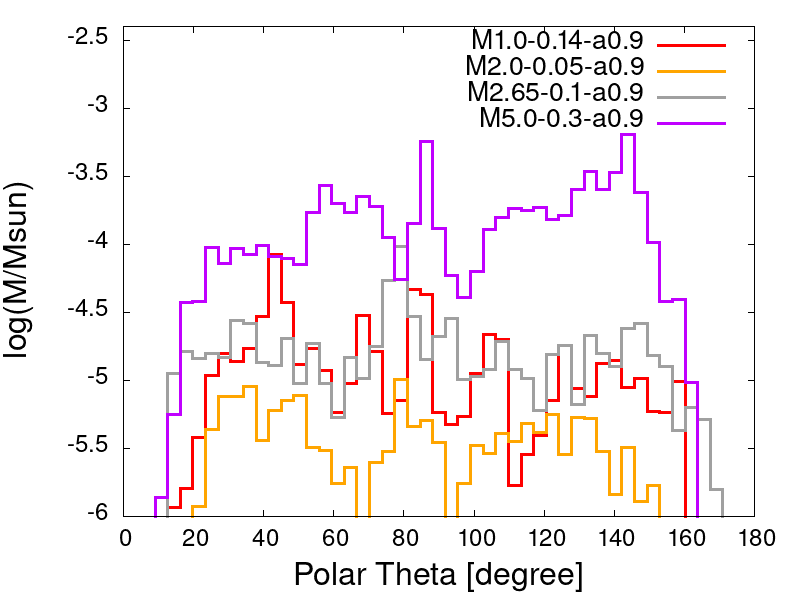}
\caption{The mass distribution versus polar angle, comparing cases with different disk and black hole mass configurations.}
\label{fig:massdistr-theta-mass}
\end{minipage}
\end{figure}

The histogram plots in Figs.~\ref{fig:massdistr-v-mass}-\ref{fig:massdistr-theta-mass} show the massive disk M5.0-0.3-0.9 produce massive ejecta over a broad range of velocities $0.01-0.3$c, but also with higher mass distribution over the lower velocities which makes the larger part of ejecta from this case slower than the others.
The broken symmetry in the outflows geometry is still visible in Fig.~\ref{fig:massdistr-theta-mass}, and overall we observe the outflows are ejected over a broad range of angles except the poles. This feature is not affected by the disk mass and/or BH mass.  

Moreover, we observe the composition of the winds is affected by the mass of the BH quite significantly. As mentioned in Sec.~\ref{sec:spin-effect}, for the LBH cases, the ejecta contains more neutron-rich material with lower average electron fraction $Y_e < 0.25$, while for the systems with stellar mass central BH, the outflow is dominated by less neutron-rich material ($Y_e > 0.25$) and therefore more lanthanide-free composition. As a result, one might expect to observe only red and IR transients from accretion disk systems with central LBH. This feature can be considered as an important signature to estimate the mass of the central postmerger remnant, and therefore an evidence for the existence of the primordial BH progenitors for such systems (\cite{Coughlin-PhRD2019}).

At this point, one might wonder how realistic these masses and geometries are compared with the remnant disks from merger simulations.  
Our selected models can present different BNS and BHNS with different parameters. 
\cite{Radice_2018} and~\cite{Coughlin-2019} derived fitting formula calculating the masses of ejecta and remnant disk for BNS systems based on numerical simulations results. Later, ~\cite{Kruger-2020} suggested a simpler version of this formula based on the same data base. Similar analytic fitting formulae were proposed by~\cite{Foucart-2012,Kawaguchi-2016,Foucart-2018} for BHNS systems. 
Using this data base and reversing this analysis to estimate the merger parameters from the black hole and accretion disk's masses, we can make the following statement about our models. 
For instance, the M5.0-0.3-a0.9 presents the remnants from a BHNS merger with mass ratio $Q \sim 3.3$, with $M_{BH}=4M_\odot$, $M_{NS}=1.2M_\odot$, and $R_{NS} = 13.2$km. Such NS can be explained by a stiff equation of state for a non-rotating spherical star such as DD2 or BHB. Our FM model gives a disk with mass of $M_{disk}=0.3M_{\odot}$ and the radius of the maximum density at $r\sim 100$km. 
Model M2.65-0.1-a0.9, resembles the remnants from a BNS merger with identical mass $M_{NS}=1.2 M_{\odot}$ and $R_{NS} = 11.7$km. Such NS can be explained by a soft equation of state for a non-rotating spherical star such as SRO and APR. The FM solution creates a disk with mass of $M_{disk}=0.1 M_{\odot}$ and the radius of the maximum density at $r \sim 44$km. 
Comparing against literature, for compact binary mergers, the remnant disk's outer radius is usually around 100-200km with maximum density around 50-60km depending on the initial parameters of the merger (\cite{Deaton-2013,Fernandez-2020}).  
Since in the FM solution, the size of the disk is scaled by the mass of the BH, this solution may create disks with larger radius.
Therefore, our disk models resembling BHNS mergers are larger and less compact compared with final disks from merger simulations. This difference can cause deviations in the mass, velocity and composition of the outflows launching from the disks with different geometries, and non-axisymmetric configurations expected for disks from mergers. 

\begin{figure}
\centering
\includegraphics[width=0.48\textwidth]{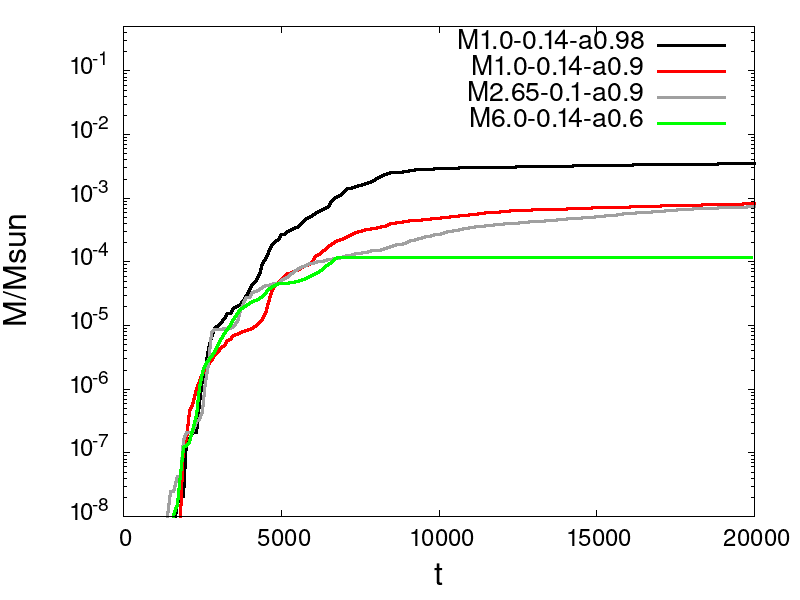}
\caption{Evolution of the cumulative outflow mass launched from the disk measured by tracers. Time is given in the code unit in this figure.   
}
\label{fig:Outflow-cum-time}
\end{figure}

Regardless of BH spin and mass configurations, we see similar pattern in the ejecta's mass evolved during time. Fig.~\ref{fig:Outflow-cum-time} shows the evolution of the cumulative ejecta's mass launched from selected models. As illustrated in this figure, the ejecta is being developed shortly after the simulations start, increases exponentially during the first half of the evolution, and then continues increasing with a slower pace for the rest of the simulation. 
This exponential growth in outflow mass at early time is consistent with the GRMHD model reported by~\cite{Fernandez-2019}, and it is a characteristic of the magnetized disk models compared with the $\alpha$-disk models. 
The detailed discussion over outflow measurements and how it is affected by the computational grid resolution and tracer setups is given in the appendix~\ref{sec:ejecta-measurment}.

\subsection{Kilonovae peak properties}

In order to investigate some important kilonovae properties based on measured outflows, we use the approximations given by~\cite{Grossman-2014} and~\cite{Dietrich-2017}. The time $t_{peak}$ at which the peak occurs, the luminosity $L_{peak}$ at this time, and the corresponding temperature $T_{peak}$ are estimated as:  

\begin{equation}
    t_{peak}  = 4.9~d \times \left( \frac{M_{ej}}{10^{-2}M_\odot} \right) ^{\frac{1}{2}} \left( \frac{\kappa}{10~cm^2g^{-1}} \right) ^{\frac{1}{2}} \left( \frac{v_{ej}}{0.1} \right)^{-\frac{1}{2}},
\end{equation}

\begin{equation} 
\begin{aligned}
    L_{peak}  = 2.5 \times 10^{40}~\text{erg/s}~\times \left( \frac{M_{ej}}{10^{-2}M_\odot} \right) ^{1-\frac{\alpha}{2}} \\
    \left( \frac{\kappa}{10~cm^2g^{-1}} \right) ^{-\frac{\alpha}{2}} \left( \frac{v_{ej}}{0.1} \right)^{-\frac{\alpha}{2}},\\
\end{aligned}
\end{equation}

\begin{equation}
\begin{aligned}
    T_{peak} = 2200~K \times \left(\frac{M_{ej}}{10^{-2}M_\odot} \right) ^{-\frac{\alpha}{8}} \\ 
    \left( \frac{\kappa}{10~cm^2g^{-1}} \right) ^{-\frac{\alpha+2}{8}} \left( \frac{v_{ej}}{0.1} \right)^{-\frac{\alpha-2}{8}},
\end{aligned}
\end{equation}

where $\kappa= 1 cm^2g^{-1}$ is the average opacity suggested by~\cite{Grossman-2014} for less opaque material produced by weak r-process, and $\alpha = 1.3$.

The results of these peak values are given in table~\ref{tab:peaks}. 
Based on these analytic fitting formulae, the majority of our models power EM transients with their peaks in a few hours after merger and the luminosity around $\sim 10^{40}-10^{41} erg/s$. 

\begin{table*}
\begin{ruledtabular}
\begin{tabular}{ l l l l l l l l}
Model &  Outflow Mass [$M_{\odot}$] & average $Y_e$ & average $v$ [c] &  $t_{peak}$[d] & $L_{peak}$ [erg/s] & $T_{peak}$ [K] \\
\hline                                         
M1.0-0.14-a0.98 & 3.5402 $\times 10^{-3}$ & 0.105 & 0.1911 & 0.66 & 1.18 $\times 10^{41}$ &  6362 \\
M1.0-0.14-a0.9 & 8.387 $\times 10^{-4}$ & 0.113 & 0.2017 & 0.32 & 7.401 $\times 10^{40}$ & 8000 \\ 
M1.0-0.14-a0.6 & 1.828 $\times 10^{-4}$ & 0.156 & 0.1745 & 0.16 & 3.952 $\times 10^{40}$  & 10370 \\
M1.0-0.14-a0.2 & 1.24 $\times 10^{-6}$  & 0.262 & 0.1430 & 0.016 & 5.15 $\times 10^{39}$ & ... \\
M1.5-0.1-a0.9 & 7.513 $\times 10^{-4}$ & 0.158 & 0.1692 & 0.33 & 6.352 $\times 10^{40}$ & 8272 \\
M2.0-0.05-a0.9 & 2.466 $\times 10^{-4}$  & 0.243 & 0.1987 & 0.17 & 4.775 $\times 10^{40}$ & 9775 \\ 
M2.65-0.1-a0.9 & 7.691 $\times 10^{-4}$ & 0.279 & 0.1693 & 0.33 & 6.407 $\times 10^{40}$ & 8240 \\
M5.0-0.3-a0.9 & 6.5585 $\times 10^{-3}$ & 0.31 & 0.1548 & 1.0 & 1.280 $\times 10^{41}$ & 5862 \\
M6.0-0.14-a0.6 & 1.211 $\times 10^{-4}$ & 0.308 & 0.1465 & 0.14 & 3.054 $\times 10^{40}$  & 11269 \\
M6.0-0.14-aR0.6 & 3.927 $\times 10^{-5}$ & 0.188 & 0.1758 & 0.07 & 2.318 $\times 10^{40}$  & 13300 \\  
\end{tabular}
\end{ruledtabular}
\caption{\label{tab:peaks}
Properties of the outflows and kilonovae peaks. 
}
\end{table*}

\subsection{Lightcurves and r-process nucleosynthesis}
\label{sec:results-kilonova}

In order to estimate the possible lightcurves powered by the ejecta from different models, we consider the same approach as~\cite{Kawaguchi-2016}. As it is shown by~\cite{Dietrich-2017} this approximation agrees well with the results given by the radiative transfer simulations for the dynamical ejecta. However, this analytic fitting overestimates the luminosity at the peak time (see Fig.8 from ~\cite{Dietrich-2017} for comparisons). Therefore, we use the~\cite{Grossman-2014} fitting formula to estimate the peak values for a better approximation.
The luminosity curve is given by

\begin{equation}
    L_{bol}(t) = (1+\theta_{ej}) \epsilon_{th} \dot{ \epsilon}_{0} M_{ej} \begin{cases} \frac{t}{t_c} \left( \frac{t}{1~d} \right)^{-\alpha}, & t \leqslant t_c \\
                                                                                        \left( \frac{t}{1~d} \right)^{-\alpha}, & t > t_c
                                                                          \end{cases}
\end{equation}

where $\dot{\epsilon}=1.58 \times 10^{10}~erg~g^{-1}~s^{-1}$ is the specific heating rate, $\epsilon_{th}$ is the efficiency of thermalization introduced by~\cite{Metzger:2010}, $0.5 < \epsilon_{th} < 1$ and $t_c$ is derived by

\begin{equation}
    t_c = \sqrt{\frac{\theta_{ej} \kappa M_{ej}}{2\phi_{ej}(v_{max}-v_{min})}},
\end{equation}
\\
with $v_{max}$ and $v_{min}$ are the maximum and minimum speed of the ejecta.

$\theta_{ej}$ and $\phi_{ej}$ are geometrical parameters of the outflows.
It was shown by~\cite{Dietrich-2017} that if we assume the ejecta as homogeneously distributed material moving with velocity $v_{ej}$ in a $\rho-z$ plane, the polar and azimuthal opening angles $\theta_{ej}$ and $\phi_{ej}$ are related to the velocity components with:

\begin{equation}
    \theta_{ej} \approx \frac{2^{4/3} v^2_{\rho} - 2^{2/3} \left(v^2_{\rho} \left(3v_z + \sqrt{9v^2_z+4v^2_{\rho}}\right) \right)^{2/3}}{\left(v^5_{\rho} \left(3v_z+\sqrt{9v^2_z+4v^2_{\rho}}\right)\right)^{1/3}},
\end{equation}

\begin{equation}
    \phi_{ej} = 4 \theta_{ej} + \frac{\pi}{2}.
\end{equation}

\begin{figure}
    \centering
    \begin{minipage}{0.48\textwidth}
    \includegraphics[width=\textwidth]{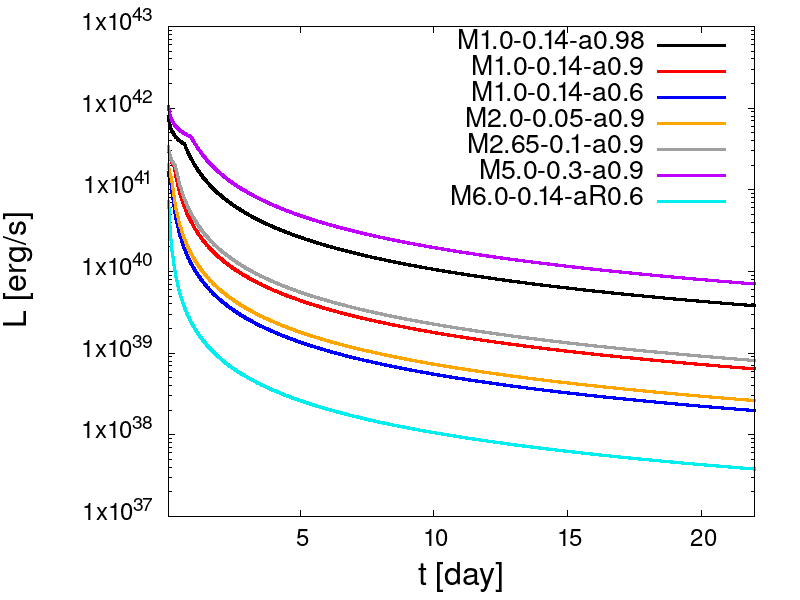}
    \caption{Luminosity versus time estimated by Kawaguchi's method for different cases.}
    \label{fig:L-vs-time}
    \end{minipage}
\end{figure}

Fig.~\ref{fig:L-vs-time} shows the predicted lighcurves from this method. As one expects, M1.0-0.14-a0.98 and M5.0-0.3-a0.9 cases with the highest mass outflows produce brighter transients. There are two pairs of models launching outflows with similar features and therefore generates lightcurves following each other closely, M2.65-0.1-a0.9 and M1.0-0.14-a0.9, and the other pair is M2.0-0.05-a0.9 and M1.0-0.14-a0.6. 
Our models suggest that although bright blue transients is possible to be observed from light BH-accretion disk systems, they may not be distinguishable from BNS mergers. Overall, the possible degeneracy make the mergers parameter estimation very challenging from kilonova observations.  

\begin{figure}
\centering
\begin{minipage}{0.48\textwidth}
\includegraphics[width=\textwidth]{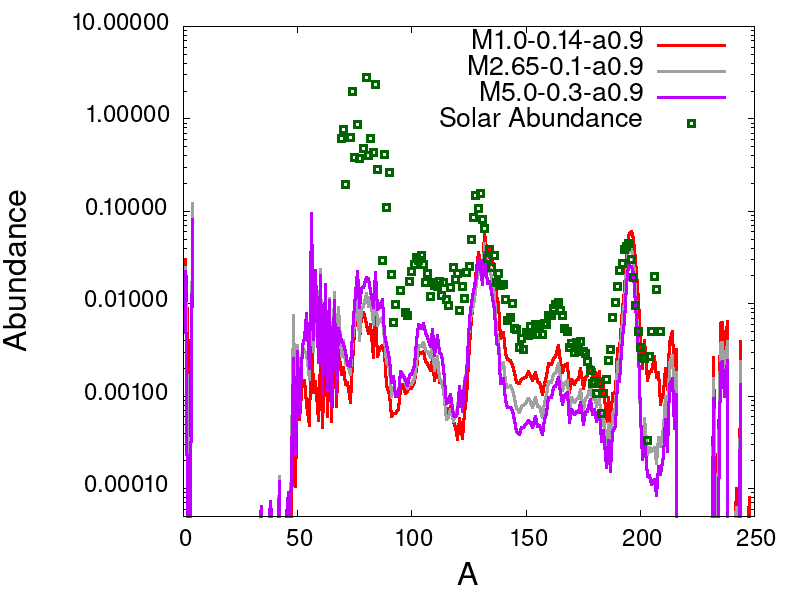}
\caption{Nuclear abundances as a function of mass number A for
each simulation, based on average of tracers sampling the outflows.
}
\label{fig:abundance}
\end{minipage}
\end{figure}

As the next post-process analysis, we have applied the r-process nucleosynthesis on the simulations outputs using the open source SkyNet code (\cite{Lippuner-2015}) to measure the nuclear abundances. The results of these simulation are illustrated in Fig.~\ref{fig:abundance} by taking the average over all the tracers and compared with the Solar system abundances. The results show the 2nd and 3rd r-process abundance peaks are resolved well in our models.
We would like to emphasise again that the composition analysis discussed in sections.~\ref{sec:spin-effect} and ~\ref{sec:mass-effect} for Figs.~\ref{fig:massdistr-v-spin} and ~\ref{fig:massdistr-v-mass} are from these skynet simulations, and these figures showing the composition of the ejecta at the onset of the r-process.

\section{Discussion}
\label{sec:discussion}

\subsection{Comparison with previous studies}
 
There are two major mechanisms driving postmerger outflows, one is neutrino heating and the other is magnetically driven turbulence.
The analytical estimation by~\cite{Perego-2014} indicated that in a post-merger accretion disk  neutrino heating timescale is short enough to drive a wind within the lifetime of the disk. 
On the other hand, in the case of a weakly magnetized accretion disk, MRI can be triggered in the disk's dynamical timescale, resulting in magnetic field amplification and turbulence. The plasma is being accreted to the BH and wind is originated from the disk, while the angular momentum is transported by magnetic turbulence effects (\cite{Balbus-2002}). 
~\cite{Tanaka:2017} has shown that the expected outflow mass from a BNS postmerger remnant disk has to be around $\sim 0.03 M_\odot$ to power emissions observed in AT2017gfo (kilonova emission observed right after GW170817) as predicted by the r-process nucleosynthesis.
They also showed that both optical and near-infrared emissions are simultaneously reproduced by the ejecta with a medium $Y_e$ of 0.25.
The recent Bayesian analysis by~\cite{Ristic:2022} constrained a mass ratio of $M_w/M_d = 2.81$ (the ratio of wind mass to dynamical ejecta mass) to reproduce the observed AT2017gfo kilonova lightcurves while it is also consistent with the observed r-process elements abundance measured in the Solar system.

While the most perfect way to study kilonovae in a numerical simulation is evolving BNS and/or BHNS systems for late inspiral, merger and postmerger phases, measuring dynamical and postmeger ejecta,
most recent studies focused only on one side of the scenario. 
In the case of disk winds, ideally we want to evolve the magnetized remnant torus in a high-resolution 3D grid for a long period ($\sim 10$ seconds), however due to the limit of computational resources, long-term 2D simulations with the $\alpha-$ viscosity prescription as the main mechanism to produce the outflows were initially preferred \citep{Fernandez-2015}.
Nonetheless, there are a few studies in the literature, which evolved magnetized remnants with neutrino cooling in fully GRMHD simulations within a multi-second timescales~\cite{Fernandez-2019,Hayashi-2021}.     
Generally, in the outflow studies, some groups considered analytical disk as the initial data configuration and then perturbed them by weak magnetic field (similar to our study in this paper) or $\alpha$ viscosity, while others deal with remnant disk from a BHNS or BNS simulations, applying Cowling approximation (frozen space-time) at $t \sim 10-15$ ms after merger.

At this point, it would be interesting to compare our results with these studies, as well as the predicted values from the kilonova observations. 

For obvious reasons, the outflow properties are highly case dependent. The wind mass reported by different groups typically varied from $10^{-5}$ to $10^{-2} M_{\odot}$. We found most of our cases produce $\sim 10^{-4}-10^{-3} M_{\odot}$, which are lower by one to two orders of magnitude compared with the predicted values for AT2017gfo observation. The outflow mass might be underestimated as a results of various reasons. The full discussion on our tracer's setup and alternative measurements is given in Appendix ~\ref{sec:ejecta-measurment}.

Comparing our models with similar cases in the literature, our results confirm the conclusion made by~\cite{Kasen-2015} for $\alpha$-viscosity disk models. As they mentioned in the case of the most mass massive disk $a0.8_{M0.3}$ (disk mass$\sim 0.3 M_{\odot}$ and BH spin $a=0.8$), the outflow mass increases by almost factor of 4 compared to their $\sim 0.1 M_{\odot}$ case, but the average velocity becomes slightly smaller.
For similar cases, we observed the massive disk, M5.0-0.3-a0.9, provides the highest outflow mass compared to the intermediate mass cases such as M2.65-0.1-0.9, but in our models the difference is quite dramatic (about one order of magnitude), while the velocity is slightly bigger. However, the outflow of their $a0.8_{M0.3}$ case is still much higher than our M5.0-0.3-a0.9 by factor of 2. 
Our M2.65-0.1-a0.9 case can be compared with HS-Therm and HS-Magn models from~\cite{Janiuk-2019}. This case is initially moderately magnetized and the comparison of this case with Table(2) from the same reference indicates that the outflows properties are scaled by the initial magnetic field strength, i. e. the outflow mass and velocity increase as the initial magnetic field increases. Also,~\cite{Janiuk-2019} found that the average electron fraction in the outflows depended on the disk magnetisation, especially at the earlier time. Though, the average electron fraction we calculated for M2.65-0.1-a0.9 case is closer to the highly magnetized case 'HS-Magn' from this reference.

Here, we measured the averaged electron fraction around $0.1<Y_e<0.3$, which is consistent with the GRMHD model in~\cite{Fernandez-2019}, and it is relatively lower than the averaged value reported by~\cite{Haddadi-2022} ($0.3 < Y_e <0.39$) for viscous disks. Our results also show that outflow mass increases sharply at early times, matching well with GRMHD model, while the viscous disks have this large growth at much later times in~\cite{Fernandez-2019}. Though, the outflow masses are still lower in our models, which can be caused by the time of evolution and differences in disk parameters and equation of state (more analysis is given in the Appendix~\ref{sec:other-outflow}).
The outflow velocity measurements agree well with the previous studies and the theoretical models' prediction for the disk winds. The velocity range is around $0.1-0.2$c and is generally higher for systems with higher spins. Regarding the geometry, the outflows launch over a wide range of angle which agrees with results reported in~\cite{Fernandez-2019} and~\cite{Fahlman2022}. However, our results show less equatorial symmetry compared with these studies.  

\subsection{Parameter estimation, binary population synthesis and spin effects}

Based on population synthesis studies, the GW from BNS mergers are more likely to be observed with EM counterparts (\cite{Mapelli-2018}).
As already discussed in the literature there are some features in the kilonova emissions, which help to distinguish the BHNS from BNS mergers and make some parameter estimations.
For instance, ~\cite{Kasen-2015} claimed that there is a rapid decline in the lightcurves from a BHNS postmerger ejecta (or equivalently, a prompt collapse for a BNS merger). Unfortunately, our current models do not allow us to investigate different possible scenarios for BNS mergers, such as creation of a magnetized and diferentially rotating HMNS with different lifetimes, which can affect the amount of ejected matter significantly (~\cite{Haas-2022}). However, even the results from~\cite{Kasen-2015} show that the light curves of models with high spin BH and HMNS with 30ms lifetime are quite similar, and this degeneracy introduces a challenge in using kilonova observations to estimate the lifetime of a HMNS remnant.

For the BHNS merger cases, two signals GW200105 and GW200115 have been detected by the LIGO-Virgo collaborations so far. 
The masses of GW200115 and GW200105 are 8.9M and 1.9M,
and 5.7M and 1.5M, respectively.
The spin of the black hole in GW200115 is not tightly constrained but it is estimated to be $\sim -0.5-0.04$ and most likely misaligned.
The dimensionless spin magnitude of the black hole in GW200105 is estimated to be $< 0.2$ and its direction is unconstrained (\cite{LIGO-2021}).
The spins of the remnant central BHs are not constrained up to certain points, but they are likely around $\sim 0.38$ and $0.43$ for GW200105 and GW200115 respectively (\cite{Chattopadhyay-2022})
However, no EM counterpart have been observed from these mergers.

The discussion given by~\cite{Steinle-2022} indicated that only BHNS with highly spinning black holes can generate massive dynamical ejecta (the difference may exceed three orders of magnitude from low spin ($\sim 0.2$) to high spin $\sim 0.9$). This high spin can be obtained by two mechanisms during a BHNS binary formation: i) inheritance due to weak core-envelope coupling of the stellar progenitors, and ii) accretion from the companion star during stable mass transfer. 
However, population synthesis studies of merger rates of BHNSs suggest that the majority of binaries will not result in observable EM counterparts. Studies by ~\cite{Drozda-2020} found that only a fraction ($\sim 20\%$) of BHNS binaries gain a high dimensionless BH spin from their stellar progenitors and produce massive ejecta during merger.

For postmerger outflows, we obtained similar trends in our results, i.e. the disk with lower spin BH generates lower mass ejecta, up to two orders of magnitude compared to high spin cases. We also concluded that the ejecta mass may become lower for the retrograde accretion.  
However the mass of the postmerger remnant disk can be another important factor for launching the ejecta, which is highly dependent on the mass ratio, NS equation of state and BH spin (\cite{Foucart-2013,Lovelace-2013}).
\footnote{Simulations done by~\cite{Lovelace-2013} show that the massive remnant disk is likely to be originated from BHNS mergers with high spin BH.} 
Overall, our models confirm that almost not much ejecta and no EM counterparts are expected to be observed from these BHNS systems, unless for BHs with high spins and lower mass ratio cases providing massive postmerger disks.

\subsection{Outflow measurements and other important physical components}

In this section we present a discussion about important numerical and physical elements which are ignored in our simulations and can affect our outflow measurements significantly.
First, it is worth mentioning that the accuracy of our outflow studies are affected by the current version of neutrino and equation of state treatments. Neutrino emissions can affect the disk's thermal and composition evolution significantly at the early time of evolution after merger (\cite{Nouri-2018,Fernandez-2019}). Therefore, a more advanced neutrino treatment such as leakage or neutrino radiation transport schemes along with composition evolution are needed to provide a better accuracy. 
Our equation of state does not include nuclei heavier than Helium,
however ~\cite{Haddadi-2022} and~\cite{Fahlman2022} showed the inclusion of these nuclei in the equation of state may cause significant changes in the outflow's mass and velocity.

In addition, the numerical scheme and grid resolution typically have impacts on the modelling of the kilonova emission.
In the study by ~\cite{Most-2019} the authors compared the dynamical ejecta properties measured from 4th-order and regular 2nd-order finite difference schemes for magnetized BNS simulations. They found the second-order scheme overestimates the amount of proton-rich shock-heated ejecta. Also, ~\cite{combi-2022} performed convergence studies on the outflows from a BNS merger simulation, and they found out the highest resolution resolve the shock-heated plasma more accurately.
The postmerger ejecta can be affected in a similar way. Fully-resolved MRI effects requires high accuracy in numerical scheme and high resolution grid, and both can impact the magnetically-driven viscous heating effects which leads to changes in the composition and other features of the disk and the outflows (see Appendix~\ref{sec:tracer-tests} about the resolution test.) 

For the sake of electromagnetic lightcurve estimations, we use here a simple fitting formula, which can be correct only up to the order of magnitude. In our simple approach we assumed a constant value for the average opcacity of the ejected matter, which in reality varies a lot depending on the geometry and composition of Lanthanide/Actinide-rich matter. In order to predict the lightcurves and spectrum of the emissions accurately, one needs to apply a radioactive transfer code on the ejecta. 
Such studies have been done recently e.g. by \cite{Wollaeger2021}.

One crucial thermal component in studying outflows is the inclusion of the r-process heating and neutrino cooling terms in the long-term hydro simulations (see the discussion and eqs. (20-21) from~\cite{Foucart2021}). 
Recent study by~\cite{Haddadi-2022} for long-term 2D viscous disk simulations indicated that adding these terms to hydro equations after one second evolution increases the outflow mass by 30\%.
Another important physics to be considered for kilonova studies is neutrino's flavour oscillations. A few groups recently studied the effects of fast flavour instability (FFI) on the disk outflows in their simulations (\cite{Li&Siegel-2021,Fernandez-2022}). Neutrino flavor oscillation is reported to have moderate (or large, in some cases), impacts on the mass ejection, average velocity, and average electron fraction.

Even with very accurate numerical evolution and measurements, there are possible ways to make our results different from the observations. 
First, the measured abundances can be largely affected by the nuclear uncertainties in the r-process models, especially at late-time emissions, $t>100$ days (\cite{Kullmann-2022}).
Second, in a realistic merger and post-merger scenario, the emissions from disk winds can be affected by the dynamical ejecta. According to the discussion by~\cite{Kasen-2015}, dynamical ejecta is generally faster and more neutron-rich and therefore more opaque, which acts like a 'Lanthanide curtain’, masking the emissions originating from the wind ejecta. They observed that for an edge-on orientation, the dynamical ejecta partially blocks the wind's optical flux by an order of magnitude.

\section{Conclusions and Summary}
\label{sec:conclusion}

We have carried out two-dimensional simulations of several black hole-accretion disk models with different initial setups using the GRMHD HARM-COOL code. These simulations include both magnetic field evolution and neutrino emission effects, as well as realistic nuclear equation of state. The initial magnetic field has been seeded in poloidal loops configuration confined within the plasma.
We evolved these models about $t \sim 200$ms, and used particle tracers to measure the outflows properties. 

We observed there is strong correlation between BH spin and the outflows properties. Generally, disks with higher mass and BH spin generate faster and more massive outflows.
We observed our models generate winds with moderate velocity (v/c $\sim 0.1-0.24$) and a broad range of electron fraction ($Y_e \sim 0.01-0.5$), which are consistent with previous studies on GRMHD postmerger accretion disk simulations. 
We have included a few cases with lower mass black holes representing BH formed from primordial black hole and neutron star mergers, and we found out such cases generate more neutron-rich ejecta in comparison with regular BHNS and BNS mergers.  

Generally, the outflow masses measured by tracers are lower compared with long viscous disk and GRMHD simulations.
We have investigated the accuracy of the outflow measurements with tracer method by altering the tracer's density threshold and extraction radius, and also grid resolution (explained in the Appendix~\ref{sec:ejecta-measurment}). We found that our measurements for ejecta's mass can be affected up to $50\%$ by this alterations.
However, we consider this method to be more accurate than the unbound matter estimation from the geodesic and Bernoulli criteria.

The general properties of postmerger ejecta derived from our simulations give us the opportunity to estimate the luminosity lightcurves of possible radioactively powered transients using analytic fitting formula. We found the luminosity peaks within the range of $\sim 10^{40}-10^{42}$ erg/s, brighter peaks for cases with higher ejecta mass, which agrees with previous studies for neutrino-driven disk wind models. 
Applying the r-process nucleosynthesis code on our results and comparing against the Solar-system abundances showed that the 2nd and the 3rd r-process abundance peaks are resolved well in our models. 

In the end, we should point out that for future studies we need to improve our numerical method by including a more sophisticated neutrino treatment such as leakage or transport schemes, as well as composition evolution to capture all the neutrino absorption/emission features. Moreover, including heavy nuclei 
in the equation of state are required to achieve a more realistic scenario with better accuracy.  

\begin{acknowledgements}
The authors thank Oleg Korobkin for helpful discussion and advice over the course of this project. This work was supported by grant No. 2019/35/B/ST9/04000 from the Polish National Science Center, Poland. We also acknowledge the support from the PL-Grid infrastructure under computational grant plgmicrophysics.  

\end{acknowledgements}

\bibliography{references}

\appendix 
\section{Outflow measurements: methods and accuracy}
\label{sec:ejecta-measurment}

In this appendix first, we investigate the alternative methods to measure the outflow mass and compare them with the results from the tracer method. In the second part, we perform several test simulations for a single case to study the effects of the resolution and tracer's setup parameters on the outflow measurements. At the end, we leave a comment regarding the evolution's time and its effect on our measurements.

\subsection{Unbound matter estimation: Geodesic and Bernoulli criteria versus tracer methods}

As mentioned, in this study we use the tracer particle technique to measure the outflows. In this method, all the particle trajectories leaving the outer boundary (or a large radius close to the outer boundary) of the computational grid during the evolution are marked as outflow winds. For the results discussed in Sec.~\ref{sec:results} we set this radius at $r=800~r_g$. The details of the tracer method's implementation in HARM-COOL are given in ~\cite{Janiuk-2019}.
The tracer particles provide the information on how different quantities such as density, temperature and electron fraction vary while tracking the particles. This information is used for the postprocess calculations of the element abundances from the r-process nucleosynthesis presented in Sec.~\ref{sec:results-kilonova}. In addition, a detailed picture of wind's geometry can be obtained from the trajectories of the outflowing particles.
Despite the fact that tracers are useful tools for outflow studies, the amount of measured outflows can be possibly underestimated if the system is not evolved long enough to provide enough time for the unbound mass to leave the grid. In fact, $t \approx 200$ms we have for these simulations can not be considered as a very long evolution to study the disk winds.
On the other hand, the evolution of the disk ejecta can be affected by several other factors including the artificial atmosphere controls over the low density regions and the tracers' initial setup parameters. 
To test the accuracy of our tracers measurements, we apply the other alternative methods from the literature to identify the unbound matter.

Referring to the discussion given by~\cite{Foucart2021}, the most accurate way to study outflows is a long-term (multi-second) 3D simulation where all the r-process heating and neutrino cooling terms are included in the source terms of the hydrodynamic evolution equations (see Eqs.(20-21) from the same reference for more details). However it is still possible to obtain a reasonable estimation for the outflow mass based on the energy criteria from a short-term simulation. We try three different criteria introduced in this paper and compare them with the results from tracers for our M2.65-0.1-a0.9 case: 1- The geodesic criterion $u_t<-1$, which is not suitable for hot disk outflows, because it ignores the thermal energy of the fluid and underestimates the unbound mass significantly. 2- The Bernoulli crtieria $hu_t < -h_{\infty}$, which includes the thermal energy but ignores the cooling effects caused by neutrino losses, and therefore overestimates the unbound mass. 3- The improved version of the Bernoulli criterion, which includes the r-process heating and neutrino losses by a simple approximation given by

\begin{equation}
    hu_t/h_{\infty}(0.9968+0.0085Y_e) < -1.
\end{equation}

Here $h$ is the enthalpy given by $h=1 + \epsilon + P/\rho$ with $\epsilon$ represent the specific internal energy, $h_{\infty}$ is the asymptotic enthalpy for $\rho \to 0$ and $\epsilon \to 0$ (for our case $h_{\infty}=1$) and $Y_e$ is the electron fraction. Generally, the improved version of the Bernoulli's criterion can be considered as a more realistic estimation to flag the unbound mass for different merger and post-merger scenarios.

The comparison for ejected mass computed from tracer method and estimation from geodesic, Bernoulli and improved Bernoulli criteria calculated at a single time snapshot in the middle of the simulation at $t \sim 100$ms for $M2.65-0.1-a0.9$ case is given in Table~\ref{tab:unbound}. These results show that the outflow mass computed from tracers is lower by about factor of four compared with the improved version of Bernoulli criterion. However, we should remind ourselves that this criterion is still considered as an approximation, and can not be taken as an accurate measurement of the unbound mass. 

\begin{table}
\begin{ruledtabular}
    \centering
    \begin{tabular}{ l l l l l}
    Model & Mass$_{trc}$ & Mass$_{GC}$ &  Mass$_{BC}$ & Mass$_{IBC}$ \\
         \hline
         M2.65-0.1-0.9 &  7.691 $\times 10^{-4}$ & 2.7316 $\times 10^{-3}$ & 3.5505 $\times 10^{-3}$ & 3.0951  $\times 10^{-3}$ \\
    \end{tabular}
    \caption{Ejected mass: tracers versus geodesic, Bernoulli and improved Bernoulli criteria (at $t \sim 100$ms snapshot).
    The volume integral is taken from $r=400~r_g$ ($\sim 1500$km) to the outer boundary.}
    \label{tab:unbound}
    \end{ruledtabular}
\end{table}

\subsection{Tracer's Setup: Grid Resolution, density threshold and extraction radius effects}
\label{sec:tracer-tests}

Investigating the effects of the computational gird resolution and tracers setup parameters, we perform several tests for our M2.65-0.1-a0.9 model, and compare their results with the standard M2.65-0.1-a0.9-Std case discussed in Sec.~\ref{sec:results}. 
This is our closest case for a BNS postmerger scenario.  
The list of tests and the measured ejecta mass and velocity are given in Table~\ref{tab:tracer-setup}. For Trac-Low-RhoMin test we alter the threshold density identifying the active tracers during the evolution and reduce this value by two orders of magnitude. The original threshold for the standard case was about $\sim 10^{-4}$ of the maximum density. Trac-1500km is designed to study the effect of the radius where we extract the information about the outflows. This radius has been set to $r=800~r_g$, which is almost 3000km for the standard case. Finally High-Res test which has the identical setup as M2.65-0.1-a0.9-Std, but with higher grid resolution 480*426 along the radial and angular directions respectively. 

\begin{figure}
\centering
\begin{minipage}{0.48\textwidth}
\includegraphics[width=\textwidth]{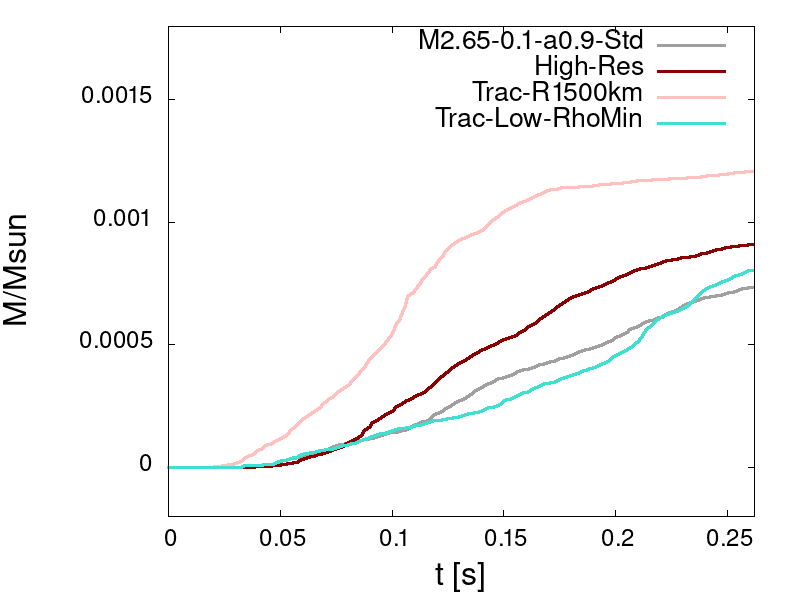}
\caption{Evolution of the cumulative outflow mass launched from
the disk for M2.65-0.1-a0.9 case with different resolution and tracer setups.}
\label{fig:Outflow-cum-M2.65-0.1-a0.9}
\end{minipage}
\quad
\begin{minipage}{0.48\textwidth}
\includegraphics[width=\textwidth]{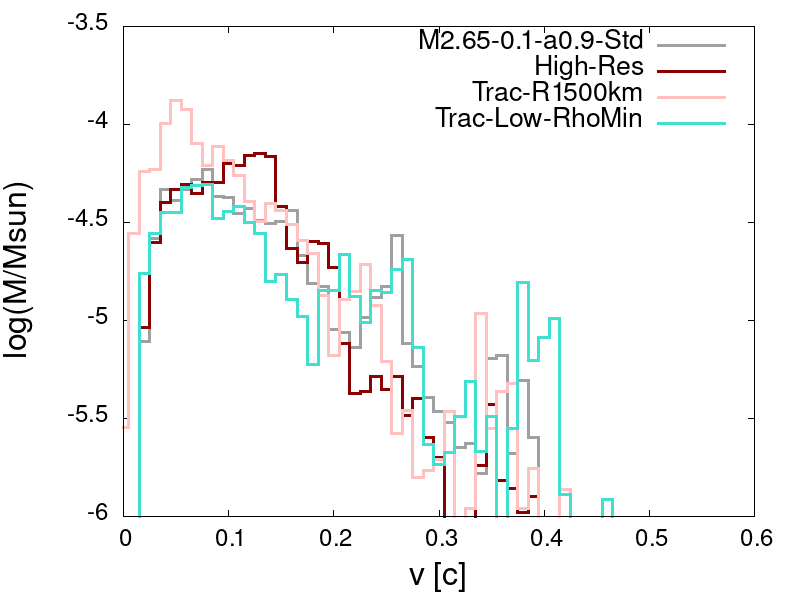}
\caption{The outflows mass distribution versus velocity for M2.65-0.1-a0.9 case with different resolution and tracer setups.}
\label{fig:massdistr_v_M2.65-0.1-a0.9}
\end{minipage}
\end{figure}

Figs.~\ref{fig:Outflow-cum-M2.65-0.1-a0.9} and~\ref{fig:massdistr_v_M2.65-0.1-a0.9} show the cumulative outflow mass versus time and the mass distribution versus velocity respectively for these tests. The simulation with lower density threshold measures slightly larger mass outflows, with larger mass distribution over the high velocity ranges. This makes the average velocity for this test significantly higher than the other cases. Extraction radius is another important parameter to be explored. Our results show that more massive and slower ejecta are measured by the smaller extraction radius. However, this mass is still lower by more than factor of two compared to the unbound mass measured by the improved Bernoulli criterion reported in Table~\ref{tab:unbound}. Specifying the extraction radius can be a challenging job, and it can affect some quantities' measurements such as velocity. In the literature, a smaller extraction radius $\sim 200-500$km has been used to measure the dynamical ejecta or the ejecta from magnetized postmerger HMNS with tracers (see for example~\cite{combi-2022} and~\cite{Haas-2022}), and each tracer has to satisfy the geodesic or Bernoulli condition to identify the unbound mass. However the disk wind is a different scenario; the recent studies by~\cite{Haddadi-2022} showed that if the r-process heating source terms are included in the evolution of the postmerger ejecta during a multi-second simulation, the ejecta's velocity need to be measured at a much larger distance. This study claimed that the ejecta is being continuously accelerated as a result of the r-process heating and it reaches to its asymptotic value around $r \sim 40,000$km.

The High-Res test shows that our measured mass can be affected by the grid resolution up to some extend. For this particular case we have the ejacta mass increased by about 12\%. Fig.~\ref{fig:massdistr_v_M2.65-0.1-a0.9} shows that the outflows with higher velocities are overestimated in the lower resolution. The standard resolution of our simulations are high enough to resolve the fastest growing MRI mode almost everywhere in the torus (see Fig.(2) from~\cite{Janiuk-2019} for a similar case). However, fully resolved MRI modes and capturing all the transport effects of turbulence requires a very high grid resolution. Even high resolution GRMHD simulations reported only a qualitative convergence in capturing all the turbulence's heating effects caused by the magnetorotational instability (\cite{Kiuchi-2015}). Obviously, higher resolution simulations provide higher accuracy for resolving the heating effects of the turbulent plasma and produce more massive outflows.

In conclusion, the results of these tests explain that our outflows measurements are affected by the resolution and the tracers setup parameters up to $50\%$, but still unable to explain the entire huge gap between the ejected mass from our model and the value predicted from GW170917 kilonova observation.

\begin{table*}
\begin{ruledtabular}
\begin{tabular}{ l l l l l l}
Model & $R_{extr}$ [km] & $\rho_{thr}$ [*$\rho_{max}$] & grid & Mass [$M_{\odot}$] & $v_{ave}$ [c] \\
\hline
M2.65-0.1-0.9-Std & 3000 & $10^{-6}$ & 384*300 & 7.691 $\times 10^{-4}$ & 0.1693 \\
Trac-Low-RhoMin & 3000 & $10^{-8}$ & 384*300 & 7.922 $\times 10^{-4}$ & 0.3190 \\ 
Trac-R1500km  & 1500 & $10^{-6}$ & 384*300 & 1.2 $\times 10^{-3}$ & 0.149 \\
High-Res & 3000 & $10^{-6}$ & 480*426 & 8.622 $\times 10^{-4}$ & 0.1853 \\
\end{tabular}
\end{ruledtabular}
\caption{\label{tab:tracer-setup} Different resolution and tracers setups for outflow measurements for M2.65-0.1-0.9 case.}
\end{table*}

\subsection{Other important factors for outflow measurements}
\label{sec:other-outflow}

One might argue that the numerical treatments can impose quantitative and qualitative effects on the outflows. Though, our results are less likely to be suffered from the numerical treatment over the atmosphere. In HARM we apply the velocity control only over low-density atmosphere with very high velocity. The disk outflow is dominated by subrelativistic plasma, and therefore, is not influenced significantly by the artificial atmosphere adjustments.

The evolution time is another key factor we should take into account. We evolved our models for about $\sim 200$ms, while the viscous timescale for a thick postmerger accretion disk with $H/R \sim 0.3$ is around $0.3$s. The evolution of about a few viscous timescales is required to capture all the magnetically and thermally driven effects for launching the outflows. 
On the other hand, the evolution time is perhaps not long enough, so not all the unbound matter have enough time to leave the grid, though the longer evolution is somehow pointless for our simulation as it is impossible to maintain the magnetic field in a long two-dimensional simulation due to the anti-dynamo theorem.
Even from thermal evolution point of view, the timescale of the evolution can be very effective based on studies by~\cite{Fernandez-2019}. They showed that a 3D GRMHD model ejects mass in two ways: one is the MHD-driven outflow at the earlier time of evolution when the torus is NDAF (neutrino-dominated accretion flow), and the second way is late-time, thermally driven wind, which occurs when the disk becomes advection-domianted (ADAF). They showed that the total amount of unbound mass ejected can reach to $0.013 M_{\odot}$, which is around 40\% of the initial disk. Half of this mass lost over the first second of the evolution and the rest has launched thereafter.

~\cite{Janiuk-2019} investigated these types of outflows in two separate models: thermally-driven winds being ejected from an initially weakly magnetized disk and magnetically-driven winds launching from strongly magnetized disk. Computing the time averaged mass loss rate over the outer boundary estimated the final mass loss to be in the range of $2-16\%$ of the initial disk. 
In comparison, our models are initially moderately magnetized (with $\beta = 50$), so they are neither purely thermally-driven nor purely MHD-driven outflows from the very beginning. 
However, the cumulative outflow mass in Fig.~\ref{fig:Outflow-cum-time} shows that the outflow mass increases in a constant and slower pace at the later time of the evolution. Assuming this mass loss rate for later times, by extrapolation, we estimate to obtain about $\sim 0.0035 M_{\odot}$ and $\sim 0.02 M_{\odot}$ mass loss after about 1 second and 9 seconds of evolution respectively; The latter would be around $20\%$ of the initial disk's mass for M2.65-0.1-a0.9 case.

\end{document}